\documentclass[10pt]{article}

\newcommand{\ben}{\begin{enumerate}}
\newcommand{\een}{\end{enumerate}}
\newcommand{\beq}{\begin{equation}}
\newcommand{\eeq}{\end{equation}}
\newcommand{\bse}{\begin{subequation}}
\newcommand{\ese}{\end{subequation}}
\newcommand{\bea}{\begin{eqnarray}}
\newcommand{\eea}{\end{eqnarray}}
\newcommand{\bc}{\begin{center}}
\newcommand{\ec}{\end{center}}

\def\DR{\rm I\kern-1.45pt\rm R}
\def\DC{\kern2pt {\hbox{\sqi I}}\kern-4.2pt\rm C}
\def\DH{\rm I\kern-1.5pt\rm H\kern-1.5pt\rm I}
\begin{document}

\begin{center}
{\Large\bf Spheroidal analysis of the generalized \\ [2mm]
MIC-Kepler system}\\[3mm]
{\bf Levon Mardoyan} \\[3mm]
International Center for Advanced Studies,\\ Yerevan State
University, \\
1, Alex Manoogian st., 375025, Yerevan, Armenia
\end{center}

\vspace{3mm}
\begin{abstract}
This paper deals with the dynamical system that generalizes the
MIC-Kepler system. It is shown that the Schr\"{o}dinger equation
for this generalized MIC-Kepler system can be separated in prolate
spheroidal coordinates. The coefficients of the interbasis
expansions between three bases (spherical, parabolic and
spheroidal) are studied in detail. It is found that the
coefficients for this expansion of the parabolic basis in terms of
the spherical basis, and vice-versa, can be expresses through the
Clebsch-Gordan coefficients for the group $SU(2)$ analytically
continued to real values of their arguments. The coefficients for
the expansions of the prolate spheroidal basis in terms of the
spherical and parabolic bases are proved to satisfy three-term
recursion relations.
\end{abstract}

\vspace{0.5cm}

\setcounter{equation}{0}

\section{Introduction}

The generalized MIC-Kepler system is described by the
equation~\cite{mard1}
\bea \frac{1}{2}\left(-i{\bf{\nabla}}- s{\bf A}\right)^2\,\psi
+\left[\frac{{s}^2}{2 r^2}-\frac{1}{r}+\frac{c_1}{r(r+z)} +
\frac{c_2}{r(r-z)}\right]\psi=E\psi, \label{schr} \eea
where $c_{1}$ and $c_{2}$ nonnegative constants, and
\begin{eqnarray}
{\bf A}=\frac{1}{r(r-z)}(y,-x,0), \qquad {\rm and} \qquad {\rm
rot}{\bf A}=\frac{{\bf r}}{r^3}. \label{field}
\end{eqnarray}
(We use the system of units for which $\hbar=m=e=c=1$.) The
monopole number $s$ satisfies the Dirac's rule of charge
quantization $s=0,\pm1/2,\pm 1,\ldots$. Each value of $s$
describes its particular generalized MIC-Kepler system. The
Schr\"{o}dinger equation (\ref{schr}) for $c_{i}=0 (i=1,2)$ and
$s\neq 0$ reduces to the Schr\"{o}dinger equation of the
MIC-Kepler system~\cite{Z, mic}. The MIC-Kepler system could be constructed
by the reduction of the four-dimensional isotropic oscillator by the use of the
so-called Kustaanheimo-Stiefel transformation both on classical and quantum
mechanical levels \cite{nt}. In the similar way, reducing the two-
and eight- dimensional isotropic oscillator, one can obtain the
two-  \cite{ntt}  and five-dimensional  \cite{iwa1} analogs of
MIC-Kepler system. An infinitely thin solenoid providing the
system by the spin $1/2$, plays the role of monopole in
two-dimensional case, whereas in the five-dimensional case this
role is performed by the $SU(2)$ Yang monopole \cite{yang},
endowing the system by the isospin. All the above-mentioned
systems have Coulomb symmetries and are solved in spherical and
parabolic coordinates both in discrete and continuous parts of
energy spectra \cite{mardoyan1, mardoyan2}. There are generalizations of
MIC-Kepler systems on three-dimensional sphere \cite{kurochkin}
and hyperboloid  \cite{np} as well. The MIC-Kepler system has been
worked out from different points of view in Refs.~\cite{mladenov,
iwa2, junker, bn, mps}.

At $s=0$, Eq.~(\ref{schr}) is
reduced to the Schr\"{o}dinger equation for generalized
Kepler-Coulomb system~\cite{KMP1}. In the case when $s=0$ and
$c_1=c_2\neq 0$, the equation (\ref{schr}) reduces to the Hartmann
system that has been used for describing axially symmetric systems
like ring-shaped molecules~\cite{hart}.

The system described by the Schr\"{o}dinger equation (\ref{schr})
is one of the Smorodinsky-Winternitz type potentials \cite{sw}.
The Smorodinsky-Winternitz type potentials where revived and
investigated in the 1990 by Evans \cite{ev}.

In Ref.~\cite{mard1} it is shown that the variables in
Schr\"{o}dinger equation (\ref{schr}) are separated in spherical
and parabolic coordinates. In this article is shown that the
variables in Eq.~(\ref{schr}) can be separated in the prolate
spheroidal coordinates also. The system of spheroidal coordinates
a natural system for investigating many problems in mathematical
physics (see Ref.~\cite{ponomar} and references cited therein). In
quantum mechanics, the spheroidal coordinates play an important
role because they are appropriate in describing the behavior of a
charged particle in the field of two Coulomb centers. The distance
$R$ between the centers is a dimensional parameter characterizing
the spheroidal coordinates. these coordinates are changed into
spherical and parabolic coordinates as $R \to 0$ and $R \to
\infty$ respectively, if the positions of one Coulomb center and
the charged particle are fixed when taking the limits. In this
sense, the spheroidal coordinates are more general then the
spherical and parabolic coordinates.

\setcounter{equation}{0}

\section{Spherical and Parabolic Bases}

For completeness, we here present the solutions of the
Schr\"{o}dinger equation (\ref{schr}) found in~\cite{mard1}.
Eq.~(\ref{schr}) in the spherical coordinates becomes
\bea
\left\{\Delta_{r\theta}+\frac{1}{4\cos^{2}\frac{\theta}{2}}\left(\frac{\partial^{2}}{\partial
\varphi^{2}}-4c_{1}\right)+\frac{1}{4\sin^{2}\frac{\theta}{2}}\left[\left(\frac{\partial}{\partial
\varphi}+2is\right)^{2}-4c_{2}\right]+2\left(E+\frac{1}{r}\right)\right\}\psi=0,
\label{schr-sp} \eea
where
\bea \Delta_{r\theta}=\frac{1}{r^{2}}\frac{\partial}{\partial
r}\left(r^{2}\frac{\partial}{\partial
r}\right)+\frac{1}{\sin\theta}\frac{\partial}{\partial\theta}\left(\sin\theta
\frac{\partial}{\partial\theta}\right). \label{delta1} \eea
The solution of Eq.~(\ref{schr-sp} has the form
\bea
\psi_{njm}^{(s)}\left(r,\theta,\varphi;\delta_{1},\delta_{2}\right)=
R_{nj}^{(s)}\left(r;\delta_{1},\delta_{2}\right)\,
Z_{jm}^{(s)}\left(\theta,\varphi;\delta_{1},\delta_{2}\right) .
\label{wv-sp} \eea
The functions
$Z_{jm}^{(s)}\left(\theta,\varphi;\delta_{1},\delta_{2}\right)$
and $R_{nj}^{(s)}\left(r;\delta_{1},\delta_{2}\right)$ are givenly
the formulae
\bea Z_{jm}^{(s)}(\theta, \varphi; \delta_{1}, \delta_{2}
)=N_{jm}(\delta_{1},
\delta_{2})\left(\cos\frac{\theta}{2}\right)^{m_1}
\left(\sin\frac{\theta}{2}\right)^{m_2}
P_{j-m_+}^{(m_2,m_1)}(\cos\theta) e^{i(m-s)\varphi}, \label{zet}
\eea
\bea R_{nj}^{(s)}(r)= C_{nj}(\delta_1, \delta_2)(2\varepsilon
r)^{j+\frac{\delta_1+ \delta_2}{2}}e^{-\varepsilon
r}F\left(-n+j+1; 2j+\delta_1+ \delta_2+2; 2\varepsilon r\right),
\label{rad}\eea
where $P_{n}^{(\alpha,\beta)}(x)$ are the Jacobi polynomials,
$F(a; c; x)$ is the confluent hypergeometric function,
$N_{jm}(\delta_{1}, \delta_{2})$ and $C_{nj}(\delta_1, \delta_2)$
are normalization constants
\bea N_{jm}(\delta_{1},
\delta_{2})=\sqrt{\frac{(2j+\delta_{1}+\delta_{2}+1)(j-m_+)!
\Gamma(j+m_+ +\delta_{1}+\delta_{2}+1)}{4\pi\Gamma(j-m_-
+\delta_{1}+1) \Gamma(j+m_- +\delta_{2}+1)}}, \label{c1} \eea
\bea C_{nj}(\delta_1, \delta_2)=
\frac{2\varepsilon^{2}}{\Gamma\left(2j+\delta_1+
\delta_2+2\right)}\sqrt{\frac{\Gamma\left(n+j+\delta_1+
\delta_2+1\right)}{(n-j-1)!}}.\label{c2}\eea
We assume that
\bea \int\limits_{0}^{\pi}\,\sin\theta\,
Z_{j'm'}^{(s)}\left(\theta,\varphi;\delta_{1},\delta_{2}\right)
Z_{jm}^{(s)}\left(\theta,\varphi;\delta_{1},\delta_{2}\right)
d\theta\,d\varphi = \delta_{jj'}\delta_{mm'},\label{norm1}\eea
\bea
\int\limits_{0}^{\infty}\,r^{2}\,R_{nj}^{(s)}\left(r;\delta_{1},\delta_{2}\right)
R_{nj}^{(s)}\left(r;\delta_{1},\delta_{2}\right) dr = \delta_{nn'}
\label{norm2}\eea
and denote by the $\varepsilon$ the following expression
\bea \varepsilon= \sqrt{-2E} = \frac{1}{n+\frac{\delta_1+
\delta_2}{2}}. \label{epsilon}\eea
The energy spectrum has the form
\bea E \equiv E_n^{(s)} = -\frac{1}{2\left(n+\frac{\delta_1+
\delta_2}{2}\right)^{2}} \label{energy}\eea
and the quantum numbers $m$ and $j$ run through the values:
$m=-j,-j+1,\dots,j-1,j$ and
\bea j = \frac{|m+s|+|m-s|}{2}, \frac{|m+s|+|m-s|}{2}+1,\dots.
\nonumber \eea
We make the following notation also $m_{\pm}=(|m+s| \pm |m-s|)/2$
and
\bea  m_1=|m-s|+\delta_{1}=\sqrt{(m-s)^2+4c_1},\qquad
m_2=|m+s|+\delta_{2}=\sqrt{(m+s)^2+4c_2}\nonumber \eea
The wave functions (\ref{wv-sp}) are the eigenfunctions of
commuting operators $\hat{M}$ and $\hat{J}_{z}$, moreover and
\bea
\hat{M}\psi_{njm}^{(s)}(r,\theta,\varphi;\delta_{1},\delta_{2})=
\left(j+\frac{\delta_{1}+\delta_{2}}{2}\right)
\left(j+\frac{\delta_{1}+\delta_{2}}{2}+1\right)
\psi_{njm}^{(s)}(r,\theta,\varphi;\delta_{1},\delta_{2}),
\label{Emm} \eea
where
\bea \hat{M}=\hat{J}^{2}+\frac{2c_{1}}{1+\cos\theta}+
\frac{2c_{2}}{1-\cos\theta}.\label{Emm1} \eea
Here $\hat{J}^{2}$ is the square of the angular momentum~\cite{Z}
\begin{equation}
\hat{\bf J}={\bf r}\times (-i{\bf\nabla}-s{\bf A}) - s\frac{{\bf r}}{r}
\label{moment}
\end{equation}
$\hat{J_{z}}=s-i\partial/\partial\varphi$ its $z$-component and
$\hat{J_{z}}\psi=m\psi$. The operator $\hat{M}$ is written in the Cartesian
coordinates as
\bea \hat{M}=-r^2\Delta +x_ix_j\frac{\partial^2}{\partial x_i \partial x_j}
+2x_i\frac{\partial}{\partial x_i}+\frac{2isr}{r-z}
\left(x\frac{\partial}{\partial y}-y\frac{\partial}{\partial x}-is-i
\frac{c_2}{s}\right)
+\frac{2c_1r}{r+z}.\label{Emm2} \eea

Let us consider the generalized MIC-Kepler system in the parabolic
coordinates coordinates $\xi,\eta \in [0, \infty), \, \varphi \in
[0, 2\pi)$, defined by the formulae
\beq x = \sqrt{\xi \eta}\,\cos\varphi, \qquad y = \sqrt{\xi
\eta}\,\sin\varphi,\qquad z = \frac{1}{2}(\xi - \eta),
\label{parabolic}\eeq
In this coordinates the differential elements of length and volume
read
\beq dl^2 = \frac{\xi + \eta}{4}\left(\frac{d\xi^2}{\xi} +
\frac{d\eta^2}{\eta}\right) + \xi \eta d\varphi^2, \qquad dV=
\frac{1}{4}(\xi + \eta)d\xi d\eta d\varphi, \label{pmetric} \eeq
while the Laplace operator looks like
\beq \Delta =\frac{4}{\xi + \eta}\left[\frac{\partial}{\partial
\xi} \left(\xi \frac{\partial}{\partial \xi}\right) +
\frac{\partial}{\partial \eta}\left(\eta \frac{\partial} {\partial
\eta}\right)\right] + \frac{1}{\xi \eta}
\frac{\partial^2}{\partial \varphi^2}. \label{laplasian} \eeq
The substitution
\beq \psi(\xi,\eta,\varphi) = \Phi_1(\xi)
\Phi_2(\eta)\,\frac{e^{i(m-s)\varphi}}{\sqrt{2\pi}}.
\label{parwave}\eeq
separates the variables in the Schr\"{o}dinger equation and we
arrive at the following system of equations
\begin{eqnarray}
\frac{d}{d \xi}\left(\xi \frac{d\Phi_1}{d \xi}\right) +
\left[\frac{E}{2}\xi - \frac{m_1^2}{4\xi} +
\frac{1}{2}\beta + \frac{1}{2}\right]\Phi_1 &=& 0, \label{eq1}\\
[3mm] \frac{d}{d \eta}\left(\eta \frac{d\Phi_2}{d \eta}\right) +
\left[\frac{E}{2}\eta - \frac{m_2^2}{4\eta}- \frac{1}{2}\beta +
\frac{1}{2}\right]\Phi_2 &=& 0, \label{eq2}
\end{eqnarray}
where $\beta$ -- is the separation constant.

These equations are analogous with the equations of the hydrogen
atom in the parabolic coordinates \cite{landau}. Thus, we get
\begin{eqnarray}
\psi_{n_1n_2m}^{(s)}(\xi,\eta,\varphi;\delta_1,\delta_2) =
\sqrt{2}\varepsilon^{2}\Phi_{n_1m_1}(\xi)
\Phi_{n_2m_2}(\eta)\,\frac{e^{i(m-s)\varphi}}{\sqrt{2\pi}},
\label{parwave1}
\end{eqnarray}
where
\begin{eqnarray}
\Phi_{n_im_i}(x) = \frac{1}{\Gamma(m_i+1)}
\sqrt{\frac{\Gamma(n_i+m_i+1)}{(n_i)!}}\,\,e^{-\frac{\varepsilon
x}{2}}\,\,(\varepsilon x)^{\frac{m_i}{2}}\,\,F(-n_i; m_i+1;
\varepsilon x). \label{parwave2}
\end{eqnarray}
Here $n_1$ and $n_2$ are nonnegative integers
\begin{eqnarray}
n_1 = -\frac{|m-s|+\delta_1+1}{2}+\frac{\beta+1}{2\varepsilon},
\qquad n_2 =
-\frac{|m+s|+\delta_2+1}{2}-\frac{\beta-1}{2\varepsilon}.
\label{parqnumb}
\end{eqnarray}
From the last relations, taking into account (\ref{energy}), we
get that the parabolic quantum numbers $n_1$ and $n_2$ are
connected with the principal quantum number $n$ as follows
\begin{eqnarray}
n= n_1 + n_2 + \frac{|m-s|+|m+s|}{2}+1. \label{parqnumb1}
\end{eqnarray}
Excluding the energy $E$ from Eqs. (\ref{eq1}) and (\ref{eq2}), we
obtain the additional integral of motion
\begin{eqnarray}
\hat{X} &=& \frac{2}{\xi + \eta}\left[\xi \frac{\partial}{\partial
\eta}\left(\eta \frac{\partial}{\partial \eta}\right) -
\eta\frac{\partial}{\partial \xi}\left(\xi
\frac{\partial}{\partial \xi}\right)\right] +
\frac{\xi-\eta}{2\xi\eta}\frac{\partial^{2}}{\partial\varphi^{2}}
- is\frac{\xi^{2}+\eta^{2}}{\xi\eta
(\xi+\eta)}\frac{\partial}{\partial\varphi}- \nonumber \\
\\
&&-s^{2}\frac{\xi-\eta}{2\xi\eta} +
\frac{2c_{1}\eta}{\xi(\xi+\eta)} -
\frac{2c_{2}\xi}{\eta(\xi+\eta)} +
\frac{\xi-\eta}{\xi+\eta}\nonumber \label{int}
\end{eqnarray}
with the eigenvalues
\begin{eqnarray}
\beta =
\varepsilon\left(n_{1}-n_{2}+\frac{|m-s|-|m+s|+\delta_{1}-\delta_{2}}{2}\right)
\label{beta}
\end{eqnarray}
and eigenfunctions
$\psi_{n_1n_2m}^{(s)}(\xi,\eta,\varphi;\delta_1,\delta_2)$, i.e.
\begin{eqnarray}
{\hat X}\psi_{n_1n_2m}^{(s)}(\xi,\eta,\varphi;\delta_1,\delta_2)=
\beta \psi_{n_1n_2m}^{(s)}(\xi,\eta,\varphi;\delta_1,\delta_2).
\label{spectr2}
\end{eqnarray}
In Cartesian coordinates, the operator $\hat{X}$ can be rewritten
as
\begin{eqnarray}
\hat{X} &=& z\left(\frac{\partial^{2}}{\partial x^{2}} +
\frac{\partial^{2}}{\partial y^{2}}\right) -
x\frac{\partial^{2}}{\partial x \partial z} -
y\frac{\partial^{2}}{\partial y \partial z}  -
is\frac{r+z}{r(r-z)}\left(x\frac{\partial}{\partial
y}-y\frac{\partial}{\partial x}\right) - \nonumber \\[2mm]
&&- \frac{\partial}{\partial z}-s^{2}\frac{r+z}{r(r-z)} +
c_1\frac{r-z}{r(r+z)} - c_2\frac{r+z}{r(r-z)} + \frac{z}{r},
\label{intc}
\end{eqnarray}
so that it immediately follows that $\hat{X}$ is connected to the
$z$-component $\hat{I_z}$ of the analog of the Runge-Lenz vector
\begin{equation}
\hat{\bf I}=\frac{1}{2}\left[(-i{\bf\nabla}-s{\bf
A})\times{\hat{\bf J}}-\hat{\bf J}\times (-i{\bf\nabla}-s{\bf
A})\right] +\frac{{\bf r}}{r} \label{runge}
\end{equation}
via
\begin{eqnarray}
\hat{X} =\hat{I_z} + c_1\frac{r-z}{r(r+z)} - c_2\frac{r+z}{r(r-z)}
\label{intc1}
\end{eqnarray}
and coincides with $\hat{I_z}$ when $c_1=c_2=0$.

\setcounter{equation}{0}

\section{Bi-Orthogonality of the Radial Wavefunctions}

We shall prove that along with the condition (\ref{norm2}) the
radial wave functions $R^{(s)}_{nj}(r;\delta_{1},\delta_{2})$
satisfy the following additional orthogonality condition
\bea
I_{jj'}=\int\limits_{0}^{\infty}\,R_{nj'}^{(s)}\left(r;\delta_{1},\delta_{2}\right)
R_{nj}^{(s)}\left(r;\delta_{1},\delta_{2}\right) dr =
\frac{2}{\left(n+\frac{\delta_{1}+\delta_{2}}{2}\right)^{3}}
\frac{\delta_{jj'}}{2j+\delta_{1}+\delta_{2}+1}. \label{bi}\eea
This new relation is used in the next Section to derive interbasis
expansions.  It can be proved as follows.

In the integral appearing in (\ref{bi}), we substitute explicit
expressions (\ref{rad}) for
$R^{(s)}_{nj}(r;\delta_{1},\delta_{2})$ and
$R^{(s)}_{nj'}(r;\delta_{1},\delta_{2})$. Then, we take the
confluent hypergeometric function in (\ref{rad}) as an finite sum
\bea F\left(-n+j+1; 2j+\delta_1+ \delta_2+2; 2\varepsilon r\right)
= \sum_{p=0}^{n-j-1}\,\frac{(-n+j+1_{)}p}
{p!(2j+\delta_{1}+\delta_{2}+1)_p}(2\varepsilon r)^p
\label{hyper}\eea
and perform the integration term by term with the help of the
formula \cite{landau}
\begin{eqnarray}
\int \limits_{0}^{\infty} e^{-\lambda x} x^\nu F(\alpha, \gamma;
kx) \, dx = \frac{\Gamma(\nu+1)}{\lambda^{\nu+1}} \, {_2F}_1
\left( \alpha, \nu+1, \gamma ; \frac{k}{\lambda}
\right).\label{intl}
\end{eqnarray}
Applying the formula
\begin{eqnarray}
{_2F}_1 \left(a, b; c; 1\right) = \frac{\Gamma(c)\Gamma(c-a-b)}
{\Gamma(c-a)\Gamma(c-b)} \label{F2l}
\end{eqnarray}
for the hypergeometric function, we obtain
\bea I_{jj'}&=&
\frac{\Gamma(j+j'+\delta_{1}+\delta_{2}+1)}{\Gamma(2j+\delta_{1}+\delta_{2}+2)}
\left[\frac{\Gamma(n+j+\delta_{1}+\delta_{2}+1))}{(n-j-1)!(n-j'-1)!
\Gamma(n+j'+\delta_{1}+\delta_{2}+1)}\right]^{1/2} \times\nonumber \\
\\
&&\times\frac{2}{\left(n+\frac{\delta_{1}+\delta_{2}}{2}\right)^{3}}\sum_{p=0}^{n-j-1}
\frac{(-n+j+1)_{p}(j+j'+\delta_{1}+\delta_{2}+1)_{p}}{p!(2j+\delta_{1}+\delta_{2}+2)_{p}}
\frac{\Gamma(n-j-p)}{\Gamma(j'-j-p+1)}. \nonumber \label{bi1}\eea
By introducing the formula~\cite{erd}
\begin{eqnarray}
\frac{\Gamma(z)}{\Gamma(z-n)} =
(-1)^n\frac{\Gamma(-z+n+1)}{\Gamma(-z+1)} \label{gamma}
\end{eqnarray}
into (\ref{bi1}), the sum over $p$ can be expressed in terms of
the $_2F_1$ Gauss hypergeometric function of argument $1$. We thus
obtain
\bea I_{jj'}&=& \frac{1}{j+j'+\delta_{1}+\delta_{2}+1}
\left[\frac{(n-j-1)!\Gamma(n+j+\delta_{1}+\delta_{2}+1))}{(n-j'-1)!
\Gamma(n+j'+\delta_{1}+\delta_{2}+1)}\right]^{1/2} \times\nonumber \\
\\
&&\times\frac{2}{\left(n+\frac{\delta_{1}+\delta_{2}}{2}\right)^{3}}
\frac{1}{\Gamma(j-j'+1)\Gamma(j'-j+1)}. \nonumber \label{bi2}\eea
Equation (\ref{bi}) then easily follows from (\ref{bi2}) since
$\left[\Gamma(j-j'+1)\Gamma(j'-j+1)\right]^{-1}=\delta_{jj'}$.

The result provided by Eq.~(\ref{bi}) generalizes the one for the
hydrogen atom~\cite{mpta}. Indeed, orthogonality properties
similar to (\ref{bi}) hold for the Kepler-Coulomb system and
harmonic oscillator in $f$-dimensional spaces $(f\geq 2)$
\cite{mpta}.

\setcounter{equation}{0}

\section{Interbasis Expansion Between Parabolic and Spherical Bases}

The connection between spherical $(r, \theta, \varphi)$ and
parabolic $(\xi, \eta, \varphi)$ coordinates is
\bea \xi = r(1+\cos\theta), \qquad \eta =  r(1+\cos\theta), \qquad
\varphi (parabolic) = \varphi (spherical). \label{conect}\eea

Now, we can write, for fixed value energy $E_{n}^{(s)}$, the
parabolic bound states (\ref{parwave1}) as a coherent quantum
mixture of the spherical bound states (\ref{wv-sp})
\bea \psi_{n_1n_2m}^{(s)}(\xi,\eta,\varphi;\delta_1,\delta_2) =
\sum_{j=m_+}^{n-1}\,W^{j}_{n_{1}n_{2}ms}\left(\delta_{1},
\delta_{2}\right)\,
\psi_{njm}^{(s)}\left(r,\theta,\varphi;\delta_{1},\delta_{2}\right).
\label{inter}\eea
By virtue of Eq.~(\ref{conect}), the left-hand side of
(\ref{inter}) can be rewritten in spherical coordinates. Then, by
substituting $\theta = 0$ in the so-obtained equation and by
taking into account that
\bea P_{n}^{(\alpha, \beta)}(1) = \frac{(\alpha +1)_n}{n!}
\label{jacobi}\eea
we get an equation that depends only on the variable $r$. Thus, we
can use the orthogonality relation (\ref{bi}) on the quantum
number $j$. This yields
\bea W^{j}_{n_{1}n_{2}ms}\left(\delta_{1}, \delta_{2}\right) =
\frac{\sqrt{\left(2j+\delta_{1}+\delta_{2}+1\right)(j-m_+)!}}
{\Gamma(m_1+1)\Gamma(2j+\delta_{1}+\delta_{2}+2)}\,E_{n_1n_2}^{jms}\,
K^{nn_1}_{jms}, \label{inter1}\eea
where
\bea E_{n_1n_2}^{jms} = \left[\frac{\Gamma\left(j-m_-
+\delta_{1}+1\right)\Gamma(n_1+m_1+1)\Gamma(n_2+m_2+1)\Gamma\left(n+j
+\delta_{1}+\delta_{2}+1\right)}{(n_1)!(n_2)!(n-j-1)!\Gamma\left(j+m_-
+\delta_{2}+1\right)\Gamma\left(j+m_+
+\delta_{1}+\delta_{2}+1\right)}\right]^{\frac{1}{2}},
\label{E}\eea
and
\bea K^{nn_1}_{jms} = \int\limits_{0}^{\infty}
e^{-x}x^{j+m_1+\delta_{1}+\delta_{2}}F\left(-n_1; m_1+1;
x\right)F\left(-n+j+1; 2j+\delta_{1}+\delta_{2}+2; x\right)dx.
\label{K}\eea
To calculate the integral $K^{nn_1}_{jms}$, it is sufficient to
write the confluent hypergeometric function $F\left(-n_1; m_1+1;
x\right)$ as a series, integrate according to (\ref{intl}) and use
the formula (\ref{F2l}) for the summation of the hypergeometric
function $_2F_1$. We thus obtain
\bea K^{nn_1}_{jms} = \frac{(n-m_+ -1
)!\Gamma(2j+\delta_{1}+\delta_{2}+2)\Gamma\left(j+m_+
+\delta_{1}+\delta_{2}+1\right)}{(j-m_+)!\Gamma\left(n+j
+\delta_{1}+\delta_{2}+1\right)}\times \nonumber \\[3mm]
\times{_3F}_2 \left\{
\begin{array}{l}
-n_1,  -j+m_+, j+m_+
+\delta_{1}+\delta_{2}+1 \\
m_1+1,   -n+m_+ +1
\end{array}
\biggr| 1 \right\}.  \label{K1}\eea
The introduction of (\ref{E}) and (\ref{K1}) into (\ref{inter1})
gives
\bea W^{j}_{n_{1}n_{2}ms}\left(\delta_{1}, \delta_{2}\right) =
\sqrt{\frac{\left(2j+\delta_{1}+\delta_{2}+1\right)\Gamma(n_1+m_1+1)
\Gamma(n_2+m_2+1)}{(n_1)!(n_2)!(n-j-1)!(j-m_+)!\Gamma\left(j+m_-
+\delta_{2}+1\right)}}\times \nonumber\\[2mm]
\times \frac{(n-m_+ -1)!}{\Gamma(m_1
+1)}\sqrt{\frac{\Gamma\left(j-m_-
+\delta_{1}+1\right)\Gamma\left(j+m_+
+\delta_{1}+\delta_{2}+1\right)}{\Gamma\left(n+j
+\delta_{1}+\delta_{2}+1\right)}}\times \\[2mm]
\times{_3F}_2 \left\{
\begin{array}{l}
-n_1,  -j+m_+, j+m_+
+\delta_{1}+\delta_{2}+1 \\
m_1+1,   -n+m_+ +1
\end{array}
\biggr| 1 \right\}. \nonumber \label{inter2}\eea

The next step is to show that the interbasis coefficients
(\ref{inter2}) are indeed a continuation on the real line of the
Clebsch-Gordan coefficients for the group $SU(2)$. It is known
that the Clebsch-Gordan coefficient
$C^{c,\gamma}_{a,\alpha;b,\beta}$ can be written as \cite{Varsh}
\begin{eqnarray}
C_{a \alpha ; b \beta}^{c\gamma} = \left[ \frac{(2c+1) (a+\alpha)!
(c+\gamma)!} {(a-\alpha)!(c-\gamma)!(a+b+c+1)!(a+b-c)!(a-b+c)!
(b-a+c)!} \right] ^{1/2} \times \nonumber \\ [1mm]
\\ [1mm]
\times (-1)^{a-\alpha} \delta_{\gamma,\alpha+\beta}
\frac{(a+b-\gamma)!(b+c-\alpha)!}{\sqrt{(b-\beta)!(b+\beta)!}}
{_3F}_2 \left\{
\begin{array}{l}
-a-b-c-1, -a+\alpha, -c+\gamma \\
-a-b+\gamma,  -b-c+\alpha  \\
\end{array}
\biggr| 1 \right\}. \label{CG1}\nonumber
\end{eqnarray}
By using the formula \cite{Bailey}
\begin{eqnarray}
{_3F}_2 \left\{
\begin{array}{l}
s, s', -N \\
t', 1-N-t  \\
\end{array}
\biggr| 1 \right\} = \frac{(t+s)_N}{(t)_N} \> {_3F}_2 \left\{
\begin{array}{l}
s, t' - s', -N \\
t', t+s  \\
\end{array}
\biggr| 1 \right\}\label{Bailey}
\end{eqnarray}
equation (\ref{CG1}) can be rewritten in the form
\begin{eqnarray}
C_{a \alpha ; b \beta}^{c\gamma} = \left[
\frac{(2c+1)(b-a+c)!(a+\alpha)!(b+\beta)!(c+\gamma)!}
{(b-\beta)!(c-\gamma) ! (a+b-c)!(a-b+c)!(a+b+c+1)!} \right] ^{1/2}
\times \nonumber \\ [1mm]
\\ [1mm]
\times
\delta_{\gamma,\alpha+\beta}\frac{(-1)^{a-\alpha}}{\sqrt{(a-\alpha)!}}
\frac{(a+b-\gamma)!}{(b-a+\gamma)!} {_3F}_2 \left\{
\begin{array}{l}
-a+\alpha, c+\gamma+1, -c+\gamma \\
\gamma-a-b,  b-a+\gamma+1  \\
\end{array}
\biggr| 1 \right\}. \label{CG2}\nonumber
\end{eqnarray}
Finally, comparing (\ref{CG2}) and (\ref{inter2}), we obtain the
representation
\bea W^{j}_{n_{1}n_{2}ms}\left(\delta_{1}, \delta_{1}\right) =
(-1)^{n_1}\,C^{j+\frac{\delta_{1}+
\delta_{2}}{2},\,\frac{m_1+m_2}{2}}_{\frac{n+m_-
+\delta_2-1}{2},\,\frac{m_2+n_2-n_1}{2};\,\frac{n-m_-
+\delta_1-1}{2},\,\frac{m_1+n_1-n_2}{2}}. \label{W}\eea
Equation (\ref{W}) proves that the coefficients for the expansion
of the parabolic basis in terms of the spherical basis are nothing
but the analytical continuation, for real values of their
arguments, of the $SU(2)$ Clebsch-Gordan coefficients.

The inverse of Eq.~(\ref{inter}), namely
\bea
\psi_{njm}^{(s)}\left(r,\theta,\varphi;\delta_{1},\delta_{2}\right)
= \sum_{n_1=0}^{n-m_+-1}\,\tilde{W}^{n_1}_{njms}\left(\delta_{1},
\delta_{2}\right)\,\psi_{n_1n_2m}^{(s)}(\xi,\eta,\varphi;\delta_1,\delta_2),
\label{inv-inter}\eea
is an immediate consequence of the orthonormality property of the
$SU(2)$ Clebsch-Gordan coefficients. The expansion coefficients in
(\ref{inv-inter}) are thus given by
\bea \tilde{W}^{n_1}_{njms}\left(\delta_{1}, \delta_{2}\right)=
(-1)^{n_1}\,C^{j+\frac{\delta_{1}+
\delta_{2}}{2},\,\frac{m_1+m_2}{2}}_{\frac{n+m_-
+\delta_2-1}{2},\,\frac{n+m_-+\delta_2-1}{2}-n_1;\,\frac{n-m_-
+\delta_1-1}{2},\,n_1+|m-s|-\frac{n-m_- -\delta_1-1}{2}}
\label{inv-W}\eea
and may be expressed in terms of the $_3F_2$ function through
(\ref{CG1}) or (\ref{CG2}).

\setcounter{equation}{0}

\section{Prolate Spheroidal Basis}

We now pass to the prolate spheroidal coordinates
\bea x = \frac{R}{2}\sqrt{(\mu^2 -1)(1-\nu^2)}\cos\varphi, \quad
y = \frac{R}{2}\sqrt{(\mu^2 -1)(1-\nu^2)}\sin\varphi, \quad
z=\frac{R}{2}(\mu\nu +1), \label{sRcoor}\eea
where $\mu \in [0;\infty)$, $\nu \in [-1;1]$, $\varphi \in
[0,2\pi)$, and $R \in [0;\infty)$. The parameter $R$ is the
interfocus distance, and in the limits where $R \to 0$ and $R \to
\infty$, the prolate spheroidal coordinates give back the
spherical coordinates and the parabolic coordinates,
respectively~\cite{ponomar, MPSTA}.

The Laplace operator in these coordinates has the form
\bea \Delta =
\frac{4}{R^2(\mu^2-\nu^2)}\left[\frac{\partial}{\partial
\mu}\left(\mu^2-1\right)\frac{\partial}{\partial \mu} +
\frac{\partial}{\partial
\nu}\left(1-\nu^2\right)\frac{\partial}{\partial \nu}\right] +
\frac{4}{R^2\left(\mu^2-1\right)\left(1-\nu^2\right)}
\frac{\partial^2}{\partial \varphi^2}. \label{Laplace}\eea
After the substitution
\beq \psi(\mu,\nu,\varphi) = \psi_1(\mu)
\psi_2(\nu)\,\frac{e^{i(m-s)\varphi}}{\sqrt{2\pi}}
\label{sRwave}\eeq
the variables in the Schr\"{o}dinger equation (\ref{schr}) are
separated
\bea
\left[\frac{d}{d\mu}\left(\mu^2-1\right)\frac{d}{d\mu}+\frac{m_1^2}{2(\mu+1)}
-\frac{m_2^2}{2(\mu-1)}+R\mu
+\frac{ER^2}{2}\left(\mu^2-1\right)\right]\psi_1 = \lambda(R)
\psi_1, \label{eq1} \\[2mm]
\left[\frac{d}{d\nu}\left(1-\nu^2\right)\frac{d}{d\nu}-\frac{m_1^2}{2(1+\nu)}
-\frac{m_2^2}{2(1-\nu)}-R\nu
+\frac{ER^2}{2}\left(1-\nu^2\right)\right]\psi_2 = -\lambda(R)
\psi_2, \label{eq2}\eea
where $\lambda(R)$ is a separation constant in prolate spheroidal
coordinates. By eliminating the the energy $E$ from
Eqs.(\ref{eq1}) and (\ref{eq2}), we produce the operator
\bea \hat{\Lambda} = \frac{1}{\mu^2-\nu^2}
\left[\left(1-\nu^2\right)\frac{\partial}{\partial
\mu}\left(\mu^2-1\right)\frac{\partial}{\partial \mu} -
\left(\mu^2-1\right)\frac{\partial}{\partial
\nu}\left(1-\nu^2\right)\frac{\partial}{\partial \nu}\right]+\nonumber\\[2mm]
+\frac{2-\mu^2-\nu^2}{\left(\mu^2-1\right)\left(1-\nu^2\right)}
\frac{\partial^2}{\partial \varphi^2}+
2s\frac{\left(\mu+\nu\right)^{2}-\left(\mu+1\right)\left(1+\nu\right)}
{\left(\mu+\nu\right)\left(\mu-1\right)\left(1-\nu\right)}
\left(s+i\frac{\partial}{\partial \varphi}\right)+\\[2mm]
+2c_1\frac{\left(\mu+\nu\right)^{2}+\left(\mu-1\right)\left(1-\nu\right)}
{\left(\mu+\nu\right)\left(\mu+1\right)\left(1+\nu\right)}+
2c_2\frac{\left(\mu+\nu\right)^{2}-\left(\mu+1\right)\left(1+\nu\right)}
{\left(\mu+\nu\right)\left(\mu-1\right)\left(1-\nu\right)} +
R\frac{\mu\nu+1}{\mu+\nu},\nonumber
\label{Lambda}\eea
the eigenvalues of which are $\lambda(R)$ and the eigenfunctions of which are
$\psi(\mu,\nu,\varphi)$. The significance of the self-adjoint operator
$\hat{\Lambda}$ can be found by switching to Cartesian coordinates. Passing
to the Cartesian coordinates in (\ref{Lambda}) and taking (\ref{Emm2}) and
(\ref{intc}) into account, we obtain
\bea \hat{\Lambda} = \hat{M} + R\hat{X}.
\label{Lambda1}\eea
Therefore,
\bea \hat{\Lambda}\psi^{(s)}_{nqm}(\mu,\nu,\varphi; R,\delta_1, \delta_2) =
\lambda_{q}(R)\psi^{(s)}_{nqm}(\mu,\nu,\varphi; R,\delta_1, \delta_2),
\label{spectr3}\eea
where index $q$ labels the eigenvalues of the operator $\hat{\Lambda}$
and varies in the range $0\leq q \leq n-m_+ -1$.

We are now ready to deal with the interbasis expansions
\bea \psi_{nqm}^{(s)}(\mu,\nu,\varphi;R,\delta_1,\delta_2) =
\sum_{j=m_+}^{n-1}\,U^{j}_{nqms}\left(R;\delta_{1},
\delta_{2}\right)\,
\psi_{njm}^{(s)}\left(r,\theta,\varphi;\delta_{1},\delta_{2}\right).
\label{inter1}\eea
\bea \psi_{nqm}^{(s)}(\mu,\nu,\varphi;R,\delta_1,\delta_2) =
\sum_{n_1=0}^{n-m_+-1}\,V^{n_1}_{nqms}\left(R;\delta_{1},
\delta_{2}\right)\,
\psi_{n_1n_2m}^{(s)}(\xi,\eta,\varphi;\delta_1,\delta_2).
\label{inter2}\eea
for the prolate spheroidal basis in terms of the spherical and parabolic bases.
(Eq.~(\ref{inter1}) was first considered by Coulson and Joseph~\cite{Coulson}
in the particular case $s=\delta_1 =\delta_2 =0$.)

First, we consider Eq.~(\ref{inter1}). Let the operator $\hat{\Lambda}$ act
on both sides of (\ref{inter1}). Then, by using Eqs.~(\ref{Lambda1}), (\ref{spectr3}),
and (\ref{Emm}) as well as the orthonormality property of the spherical basis,
we find that
\bea \left[\lambda_{q}(R)-\left(j+\frac{\delta_1+\delta_2}{2}\right)
\left(j+\frac{\delta_1+\delta_2}{2}+1\right)\right]U^{j}_{nqms}(R;\delta_1,\delta_2) =
R\sum_{j'=m_+}^{n-1}U^{j'}_{nqms}(R;\delta_1,\delta_2)(\hat{X})_{jj'},
\label{inter11}\eea
where
\bea (\hat{X})_{jj'} = \int\,
\psi_{njm}^{(s)*}\left(r,\theta,\varphi;\delta_{1},\delta_{2}\right)
\hat{X}\psi_{njm}^{(s)}\left(r,\theta,\varphi;\delta_{1},\delta_{2}\right)dV.
\label{inter12}\eea
The calculation of the matrix element $(\hat{X})_{jj'}$ can be done by
expanding the basis in (\ref{inter12}) in terms of the parabolic wavefunctions
[see Eq.~(\ref{inv-inter})]and by making use of the eigenvalue equation for
$\hat{X}$ [see Eq.~(\ref{spectr2})]. This leads to
\bea (\hat{X})_{jj'} = \frac{2}{2n+\delta_{1}+\delta_{2}}
\sum_{n_1=0}^{n-m_+ -1}\,\left(2n_1-n+|m-s|+\frac{\delta_1+\delta_2}{2}+1\right)
\tilde{W}^{n_1}_{njm} \tilde{W}^{n_1}_{nj'm}.
\label{inter13}\eea
Then, by using Eq.~(\ref{inv-W}) together with the recursion relation~\cite{Varsh}
\begin{eqnarray}
C_{a\alpha;b\beta}^{c\gamma} =
-\left[ \frac{4c^2(2c+1)(2c-1)}{(c+\gamma)(c-\gamma)(b-a+c)(a-b+c)
(a+b-c+1)(a+b+c+1)} \right]^{1/2} \times \nonumber \\ [3mm]
\times \Biggl\{\left[\frac{(c-\gamma-1)(c+\gamma-1)(b-a+c-1)(a-b+c-1)(a+b-c+2)
(a+b+c)}{4(c-1)^2 (2c-3)(2c-1)} \right]^{1/2} \times\\ [3mm]
\times C_{a\alpha;b\beta}^{c-2,\gamma}
-\frac{(\alpha-\beta)c(c-1) - \gamma a(a+1) + \gamma b(b+1)}
{2c(c-1)} C_{a\alpha;b\beta}^{c-1,\gamma}\Biggr\}, \nonumber \label{inter14}
\end{eqnarray}
and the orthonormality condition
\begin{eqnarray}
\sum_{\alpha+\beta=\gamma} C_{a\alpha;b\beta}^{c \gamma}
C_{a\alpha;b\beta}^{c'\gamma'} = \delta_{c'c} \delta_{\gamma' \gamma}
\label{inter15}
\end{eqnarray}
we find that $(\hat{X})_{jj'}$ is given by
\begin{eqnarray}
(\hat{X})_{jj'} = -\frac{2}{2n+\delta_{1}+\delta_{2}}
\left(A^{j+1}_{nm}\delta_{j',j+1}+A^{j}_{nm}\delta_{j',j-1}\right)+
\frac{(m_1+m_2)(m_1-m_2)}{(2j+\delta_{1}+\delta_{2})(2j+\delta_{1}+\delta_{2}+2)}
\delta_{j',j},
\label{inter16}
\end{eqnarray}
where
\begin{eqnarray}
A^{j}_{nm} = \left[\frac{(j-m_+)(j+m_+ +\delta_{1}+\delta_{2})
(j-m_- +\delta_1)(j+m_- +\delta_2)(n-j)(n+j+\delta_{1}+\delta_{2})}
{\left(j+\frac{\delta_{1}+\delta_{2}}{2}\right)^2
(2j+\delta_{1}+\delta_{2}-1)(2j+\delta_{1}+\delta_{2}+1)}\right]^{1/2}.
\label{inter17}
\end{eqnarray}
Now by introducing (\ref{inter16}) into (\ref{inter11}), we get the following
three-term recursion relation for the coefficient $U^{j}_{nqms}$
\bea
&&\left[\lambda_{q}(R)-\left(j+\frac{\delta_1+\delta_2}{2}\right)
\left(j+\frac{\delta_1+\delta_2}{2}+1\right)-
\frac{R(m_1+m_2)(m_1-m_2)}{(2j+\delta_{1}+\delta_{2})(2j+\delta_{1}+\delta_{2}+2)}
\right]U^{j}_{nqms}(R;\delta_1,\delta_2) + \nonumber \\[3mm]
&&+ \frac{2R}{2n+\delta_{1}+\delta_{2}}
\left[A^{j+1}_{nm}U^{j+1}_{nqms}(R;\delta_1,\delta_2)+
A^{j}_{nm}U^{j-1}_{nqms}(R;\delta_1,\delta_2)\right]=0.
\label{inter18}\eea
The recursion relation (\ref{inter18}) provides us with a system $n-m_+$ linear
homogeneous equations which can be solved by taking into account the
normalization condition
\bea \sum_{j-m_+}^{n-1}\left|U^{j}_{nqms}(R;\delta_1,\delta_2)\right|^2 =1.
\label{inter19}\eea
Th eigenvalues $\lambda_{q}(R)$ of the operator $\hat{\Lambda}$ then follow from the
vanishing of the determinant for the latter system.

Second, let us concentrate on the expansion (\ref{inter2}) of the prolate spheroidal
basis in terms of the parabolic basis. By employing technique similar to the one used
for deriving Eq.~(\ref{inter11}), we get
\bea \left[\lambda_{q}(R)-\frac{2R}{2n+\delta_{1}+\delta_{2}}
\left(n_1-n_2+\frac{m_1-m_2}{2}\right)\right]V^{n_1}_{nqms}(R;\delta_1,\delta_2) =
\sum_{n_1'=0}^{n-m_+-1}V^{n_1'}_{nqms}(R;\delta_1,\delta_2)(\hat{M})_{n_1n_1'},
\label{inter21}\eea
where
\bea (\hat{M})_{n_1n_1'} = \int\,
\psi_{n_1n_2m}^{(s)*}\left(\xi,\eta,\varphi;\delta_{1},\delta_{2}\right)
\hat{M}\psi_{n_1'n_2'm}^{(s)}\left(\xi,\eta,\varphi;\delta_{1},\delta_{2}\right)
dV.
\label{inter22}\eea
The matrix elements $(\hat{M})_{n_1n_1'}$ can be calculated in the
same way as $(\hat{X})_{jj'}$ except that now we must use the
relation~\cite{Kibler}
\begin{eqnarray}
\left[\,c(c+1) - a(a+1) - b(b+1) - 2\alpha\beta\right]\,
C^{c,\gamma}_{a, \alpha; b, \beta} = \nonumber \\ [2mm]
= \sqrt{(a+\alpha)(a-\alpha+1)(b-\beta)(b+\beta+1)}\,
C^{c,\gamma}_{a, \alpha - 1; b, \beta + 1} +
\\ [2mm]
+ \sqrt{(a-\alpha)(a+\alpha+1)(b+\beta)(b-\beta+1)}\,
C^{c,\gamma}_{a, \alpha + 1; b, \beta - 1},   \nonumber
\label{inter23}
\end{eqnarray}
and the orthonormality condition
\begin{eqnarray}
\sum_{c=|\gamma|}^{a+b} C_{a\alpha ;b\beta }^{c\gamma}
C_{a\alpha';b\beta'}^{c\gamma}
= \delta_{\alpha\alpha'} \delta_{\beta\beta'}
\label{inter24}
\end{eqnarray}
permit deriving the formula for the matrix element $(\hat{M})_{n_1n_1'}$
\bea (\hat{M})_{n_1n_1'} = \biggl[(n_1+1)(n_2+m_-) +
(n-n_1+\delta_2)(n_1+|m-s|+\delta_2) +\frac{1}{4}(\delta_1-\delta_2)(\delta_1-\delta_2-2)
 + \nonumber \\[2mm]
+m_-(m_+ +\delta_2)\biggr]\,
\delta_{n_1'n_1} - \sqrt{n_2(n_1+1)(n_1+|m-s|+\delta_1+1)
(n_2+|m-s|+\delta_2)}\delta_{n_1',n_1+1}-\\[2mm]
- \sqrt{n_1(n_2+1)(n_1+|m-s|+\delta_1+1)
(n_2+|m-s|+\delta_2+1)}\delta_{n_1',n_1-1}.\nonumber
\label{inter25}\eea
Finally, the introduction of (\ref{inter25}) into (\ref{inter21}) leads to the
three-term recursion relation
\bea
\biggl[(n_1+1)(n_2+m_-) +
(n-n_1+\delta_2)(n_1+|m-s|+\delta_2) +\frac{1}{4}(\delta_1-\delta_2)(\delta_1-\delta_2-2)
 + \nonumber \\[2mm]
+m_-(m_+ +\delta_2)+\frac{2R}{2n+\delta_{1}+\delta_{2}}
\left(n_1-n_2+\frac{m_1-m_2}{2}\right)-\lambda_{q}(R)\biggr]\,
V^{n_1}_{nqms}(R;\delta_1,\delta_2)=
\nonumber \\ [2mm]
\sqrt{n_2(n_1+1)(n_1+|m-s|+\delta_1+1)
(n_2+|m-s|+\delta_2)}\,V^{n_1+1}_{nqms}(R;\delta_1,\delta_2)+ \\[2mm]
+\sqrt{n_1(n_2+1)(n_1+|m-s|+\delta_1+1)
(n_2+|m-s|+\delta_2+1)}\,V^{n_1-1}_{nqms}(R;\delta_1,\delta_2)\nonumber
\label{inter26}\eea
for the expansion coefficients $V^{n_1}_{nqms}(R;\delta_1,\delta_2)$.
This relation can be iterated by taking account of the normalization condition
\bea
\sum_{n_1=0}^{n-m_+-1}\left|V^{n_1}_{nqms}(R;\delta_1,\delta_2)\right|^{2}=1.
\label{inter27}\eea
Here again, the eigenvalues $\lambda_{q}(R)$ may be obtained by solving a system of
$n-m_+$ linear homogeneous equations.

It should be mentioned, that the formulae (\ref{W}) and (\ref{inv-W}) and
three-term recursion relations (\ref{inter18}) and (\ref{inter18})
generalize the analogical results for the following systems:
\begin{itemize}
\item Hydrogen atom~\cite{Park,Tarter,APTA,MPSTA,Coulson,MPSTA2}, when
$s=\delta_1=\delta_2=0$.
\item Generalized Kepler-Coulomb system~\cite{KMP1}, when $s=0$,
$\delta_1 \neq \delta_2 \neq 0$.
\item Hartmann system~\cite{LPSTA}, when $s=0$,
$\delta_1 = \delta_2 \neq 0$.
\item Charge-dyon system~\cite{mardoyan1}, when $s \neq 0$,
$\delta_1 = \delta_2 = 0$.
\end{itemize}

Finally, it should be noted that the following four limits
\begin{eqnarray}
&&\lim_{R \to 0} U_{jnqms}^{j}(R;\delta_1,\delta_2) = \delta_{j q}, \qquad
\lim_{R \to \infty} U_{nqms}^{j}(R;\delta_1,\delta_2) =
 W_{n_1n_2ms}^{j}(\delta_1,\delta_2), \\ [3mm]
&&\lim_{R \to \infty} V_{nqms}^{n_1}(R;\delta_1,\delta_2) = \delta_{n_1 q}, \qquad
\lim_{R \to 0} V_{nqms}^{n_1}(R;\delta_1,\delta_2) =
{\tilde W}_{njms}^{n_1}(\delta_1,\delta_2)
\label{inter28}
\end{eqnarray}
furnish a useful means for checking the calculations presented in
Sections 4 and 5.

\vspace{5mm}

{\large Acknowledgements.} I would like to thank Dr. Armen Nersessian for useful
discussions. The work is carried out with the support of ANSEF No: PS81 grant.

\end{document}